\documentstyle[floats,prd,aps]{revtex}
\begin{document}
\draft

\title{Calculation of gravitational wave forms
from black hole collisions and disk collapse: Applying perturbation
theory to numerical spacetimes}

\author{Andrew M. Abrahams\cite{abrahams},
Stuart L. Shapiro\cite{shapiro}, and
Saul A. Teukolsky\cite{teukolsky}}

\address{Center for Radiophysics and Space Research,
          Cornell University, Ithaca, NY 14853}

\date{\today}
\twocolumn[
\maketitle

\begin{abstract}
\widetext
Many simulations of gravitational collapse to black holes become inaccurate
before the total emitted gravitational radiation can be determined.
The main difficulty is that a significant component of the radiation
is still in the near-zone, strong field region
at the time the simulation breaks down.
We show how to calculate the emitted waveform by matching the
numerical simulation to a perturbation solution
when the final state of the system approaches a Schwarzschild black hole.
We apply the technique to two scenarios: the head-on collision of
two black holes, and the collapse of a disk to a black hole.
This is the first reasonably accurate calculation of the radiation
generated from colliding black holes that form from matter collapse.
\end{abstract}
\pacs{04.70.-s, 04.30.-w, 04.25.Dm, 04.25.Nx}
]
\narrowtext
\section{Introduction}
\label{sec:intro}

One of the most significant goals of numerical
relativity is to calculate the gravitational wave forms
from realistic, astrophysically interesting
systems.  Until now, obtaining a good estimate
of the asymptotic wave form has required long numerical evolutions
so that the emitted waves can be propagated
far from the source. There, either
radiation extraction techniques can be used, or
the solution can be matched onto a null-cone integration.
Unfortunately, long evolutions are difficult for
a number of reasons. These include
throat stretching when black holes form, numerical instabilities
associated with curvilinear coordinate systems, and
the effects of approximate outer boundary conditions.
Moreover, evolutions
using certain time-slicing conditions do not allow simulations such
as collisions
to progress far enough for the black holes
to completely merge; rather they asymptote to a finite separation.
At this time, no numerical algorithm is known that
deals effectively with all these problems in the
context of multidimensional, radiating
spacetimes.

Recently however, an important idea introduced by
Price and Pullin\cite{price_pullin94}
provides a possible way to circumvent some of these problems
in certain circumstances.
Price and Pullin considered Misner initial data representing
two black holes at a moment of time-symmetry.  They realized
that when the two holes were sufficiently close, the system
could be treated as a single, perturbed black hole.  By applying
gauge invariant perturbation theory, they calculated this
perturbation and evolved it using the Zerilli equation.
This allowed them to compute asymptotic wave forms and
emitted energies.  Remarkably, for fairly small separations,
the energies and wave forms agreed well with the results
of fully relativistic numerical simulations~\cite{anninos_etal93}.
In a subsequent study, Abrahams and Cook\cite{abrahams_cook94}
extended this technique and applied it to initial
data representing boosted black holes with a
common apparent horizon.
The radiated energies computed from these initial data
sets agreed quite well with those from the fully
relativistic simulations for cases in which the initial
separations were large.

In this paper, we apply the same technique to numerically
evolved spacetimes that at late times can be approximated
as a single perturbed black hole.   The perturbation
of the black hole is computed on a spatial slice from the numerical
metric and extrinsic curvature using gauge-invariant
perturbation theory.  Wave forms are determined by
evolving this perturbation to infinity using the Zerilli
equation.

The important difference between this method and
standard finite-radius radiation extraction techniques
(cf. \cite{ae90}) is the use of the horizon as an
inner boundary at which radiation is purely ingoing.
Once the horizon forms, and the spacetime
settles down to a quasi-static state, data from the
numerical simulation is no longer required.
The simulation can be terminated at a relatively
early epoch; it is not necessary to propagate
the radiation pulse out to the weak-field regime.
Another advantage of the perturbation method is that
the effects of backscatter of the waves
off the black hole curvature are automatically taken
into account via the Zerilli equation integration
of the perturbation to large radii.
These effects can only be incorporated approximately
with standard extraction methods in which an integration
over a timelike cylinder is used to separate off
near-zone effects.
It should be emphasized that the perturbation
approach requires the formation of a black hole
during the numerical simulation. The black hole horizon cuts off
the evolution inside the black hole from influencing the waves outside.
The method would be inadequate
for, say, an oscillating neutron star spacetime, where there is no horizon.

In Sec.~\ref{sec:methods} we briefly describe the numerical
relativity code and detail the implementation of the
gauge-invariant black hole perturbation method used to
compute wave forms.  In Sec.~\ref{sec:results} we
demonstrate the application of this method to calculations
of black hole collisions and disk collapse.

\newpage
\section{Methodology}
\label{sec:methods}

The numerical spacetimes representing colliding black holes and
collapsing disks are generated using a code
that solves the field equations of general relativity coupled
to a collisionless-particle matter source.
Full details of the equations solved and the numerical
method can be found in Ref.~\cite{st92,ast94}
The code employs the maximal time slicing condition and
the quasi-isotropic spatial gauge, and applies to nonrotating spacetimes
in axisymmetry.
The metric, expressed in spherical polar coordinates,
takes the form
\begin{eqnarray}\label{eq:metric}
ds^2=-\alpha^2 dt^2+\phi^4 e^{2 \eta/3} (dr+\beta^r dt)^2+ \nonumber \\
\phi^4 e^{2 \eta/3}
r^2(d\theta+\beta^\theta dt)^2
+\phi^4 e^{-4 \eta/3} r^2\sin^2\theta d\varphi^2.
\end{eqnarray}
The code solves the Hamiltonian constraint to compute $\phi e^{-\eta/3}$,
solves the momentum constraints for the components
$K^r_r$ and $K^\phi_\phi$ of the extrinsic curvature, and
evolves the variables $\eta$ and $K^r_\theta$.  The kinematical
variables $\alpha$ and $\beta^i$ are determined from the maximal
slicing condition and spatial gauge conditions respectively.
In Ref.~\cite{st92} the code was used to study head-on collisions
and in Ref.~\cite{ast94} it was used to study disk collapse.

A thorough discussion of the perturbed black hole approximation
will be published elsewhere~\cite{abrahams_price94}.
Here we sketch the general idea and list the equations relevant
to the current application.
The theory of gauge invariant perturbations of Schwarzschild
was first developed by Moncrief\cite{moncrief74}.  Moncrief
showed how to use the Regge-Wheeler multipole amplitudes
and their radial derivatives to construct unconstrained even and odd-parity
functions that are invariant under infinitesimal coordinate
transformations about the Schwarzschild background.
These functions satisfy the Zerilli or Regge-Wheeler equation for
the appropriate $\ell$-mode.
For our approach, which involves extraction of the
perturbation data on a spatial slice, the background
Schwarzschild time-derivative of the gauge invariant function
is also required in order to pose initial data for integration
of the second-order Zerilli equation
(see~\cite{moncrief74,abrahams_cook94,abrahams_price94}.

{}From the viewpoint of our approximation,
the nonspherical part of the three-metric, the shift, and the
extrinsic curvature are treated as perturbations
about a static Schwarzschild black hole in the
(approximate) exterior region  $r \ge M/2$, where $M$ is the
the ADM mass of the slice.  In quasi-isotropic gauge,
the static Schwarzschild metric takes the standard
isotropic-coordinate form ($\beta_i=0$, $\eta=0$, $\phi=1+M/2r$, $K_{ij}=0$).
To compute the perturbation using standard gauge-invariant
formalism, first we make the background coordinate
transformation of the  numerically
computed metric and extrinsic curvature on a given spatial slice
from isotropic to Schwarzschild radial coordinates
$r \rightarrow R$, where $R=r(1+M/2r)^2$.  Then
we compute the the following Regge-Wheeler multipole
amplitudes by integrating over coordinate two-spheres at constant $R$,
\begin{eqnarray}
H_2^{\ell}&=& - 2 \pi {r^2 \over R^2}\int_{-1}^1 dx Y_{l0} \phi^4 e^{2\eta/3},
\\
h_1^{\ell} &=&  0,     \\
G^{\ell} &=& - 2 \pi {r^2 \over R^2} \int_{-1}^1 dx \sqrt{1-x^2}
{\partial^2 Y_{\ell 0} \over \partial x^2} \phi^4(e^{2 \eta/3}-e^{-4 \eta/3}),
\\
K^{\ell} &=& {\ell (\ell+1) \over2} G^{\ell} - \pi {r^2 \over R^2}
\int_{-1} ^1 dx Y_{\ell 0} \phi^4(e^{2 \eta/3}+e^{-4 \eta/3}) ,
\end{eqnarray}
where $x \equiv \cos \theta$.  Note that we have
specialized to axisymmetry, so the azimuthal
quantum number $m=0$.
We also compute the normal Lie-derivatives of the multipole
amplitudes:
\begin{eqnarray}
{\cal L}_{\bf n} H_2^{\ell} = 4 \pi {r^2 \over R^2}\int_{-1}^1 dx Y_{l0} \phi^4
e^{2\eta/3} K^r_r, \\
{\cal L}_{\bf n} h_1^{\ell} =  2 \pi \int_{-1}^1 dx
\sqrt{1-x^2}{\partial Y_{l0} \over \partial x} \phi^4 e^{2\eta/3} K^r_\theta,
\\
{\cal L}_{\bf n} G^{\ell} = 4 \pi {r^2 \over R^2}  \int_{-1}^1 dx \sqrt{1-x^2}
{\partial^2 Y_{\ell 0} \over \partial x^2}  \phi^4
\left [-e^{2 \eta/3} \right. \nonumber \\ \left.  (K^r_r+K^\phi_\phi)
-e^{-4 \eta/3} K^\phi_\phi \right ] , \\
{\cal L}_{\bf n} K^{\ell} = {\ell (\ell+1) \over2} {\cal L}_{\bf n}
G^{\ell} + 2 \pi {r^2 \over R^2}
\int_{-1} ^1 dx Y_{\ell 0} \nonumber \\ \phi^4  [-e^{2 \eta/3}
(K^r_r+K^\phi_\phi)+e^{-4 \eta/3}K^\phi_\phi] ,
\end{eqnarray}

{}From these we form the gauge-invariant function $Q_{\ell}$\cite{moncrief74},
\begin{eqnarray}\label{eq:gi}
Q_{\ell} = \left[2 (\ell-1)(\ell+2) \over \ell (\ell+1) \right]^{1/2}
{ R \over \Lambda} \left( 2N^2 H_2^{\ell}- \right. \nonumber \\ \left.
2N^3 {\partial \over \partial R} (RK^{\ell}/N)+
\ell(\ell+1)\left [K^{\ell}+ \right. \right. \nonumber \\ \left. \left.
N^2R{\partial G^{\ell} \over \partial R} - 2{N^2 h_1^{\ell} \over R}
\right ] \right)
\end{eqnarray}
where $N^2 \equiv 1-2M/r$ and $\Lambda= (\ell-1)(\ell+2)+6M/R$.
We also compute its Schwarzschild time-derivative $\dot Q_{\ell}= N {\cal
L}_{\bf n}Q_{\ell}$,
where ${\cal L}_{\bf n} Q_{\ell}$ is constructed by replacing the amplitudes
appearing in Eq.~(\ref{eq:gi}) with their normal Lie-derivatives.
Note that  it is valid to compute the time derivative using $N{\cal L}_{\bf
n}$:
the difference between these two operators is proportional to
the Lie-derivative along the shift, ${\cal L}_\beta $, which is
higher order.  The two functions $Q_{\ell}$ and
$\dot Q_{\ell}$ provide initial data for
integration of the Zerilli equation,
\begin{equation}\label{eq:zerilli}
{\partial ^2 \over \partial t^2} Q_{\ell}
-{\partial ^2 \over \partial r_*^2} Q_{\ell}
+V_{\ell}(R) Q_{\ell} = 0
\end{equation}
where $r_* \equiv R+2M \ln(R/2M-1)$ is the tortoise coordinate
and the Zerilli potential is given by
\begin{eqnarray}
V_{\ell}(R) = N^2 \left[{1\over\Lambda^2} \left (
{72M^2\over R^5}- \right. \right. \nonumber \\
\left. \left. {12M\over R^3}(\ell-1)(\ell+2)(1-3M/R)\right)
+ \right. \nonumber \\ \left. {\ell(\ell-1)(\ell+1)(\ell+2) \over
R^2 \Lambda } \right].
\end{eqnarray}
Numerically, the initial data $(Q_{\ell}(R),\dot Q_{\ell}(R))$ is interpolated
onto a fine mesh ranging from $r_*=-500$
to $r_*=1000$ and the Zerilli equation is integrated using a 2nd-order accurate
leapfrog scheme until the perturbation has all been propagated far
away from the peak of the potential, $|r_*| \rightarrow \infty$.
At large radius the function approaches the
even-parity gravitational wave amplitude
\begin{equation}
r h_+/\sin^2 \theta=\sqrt{15 \over 64 \pi } Q_2(t;r).
\end{equation}

\section{Numerical results}
\label{sec:results}

Applying the above equations
to numerically generated spacetimes is
have criteria for when we expect the method to give reliable results.
\begin{figure}
\label{fig:eoft}
\special{hscale=0.45 vscale = 0.45 hoffset = -0.2 voffset = -4.
psfile=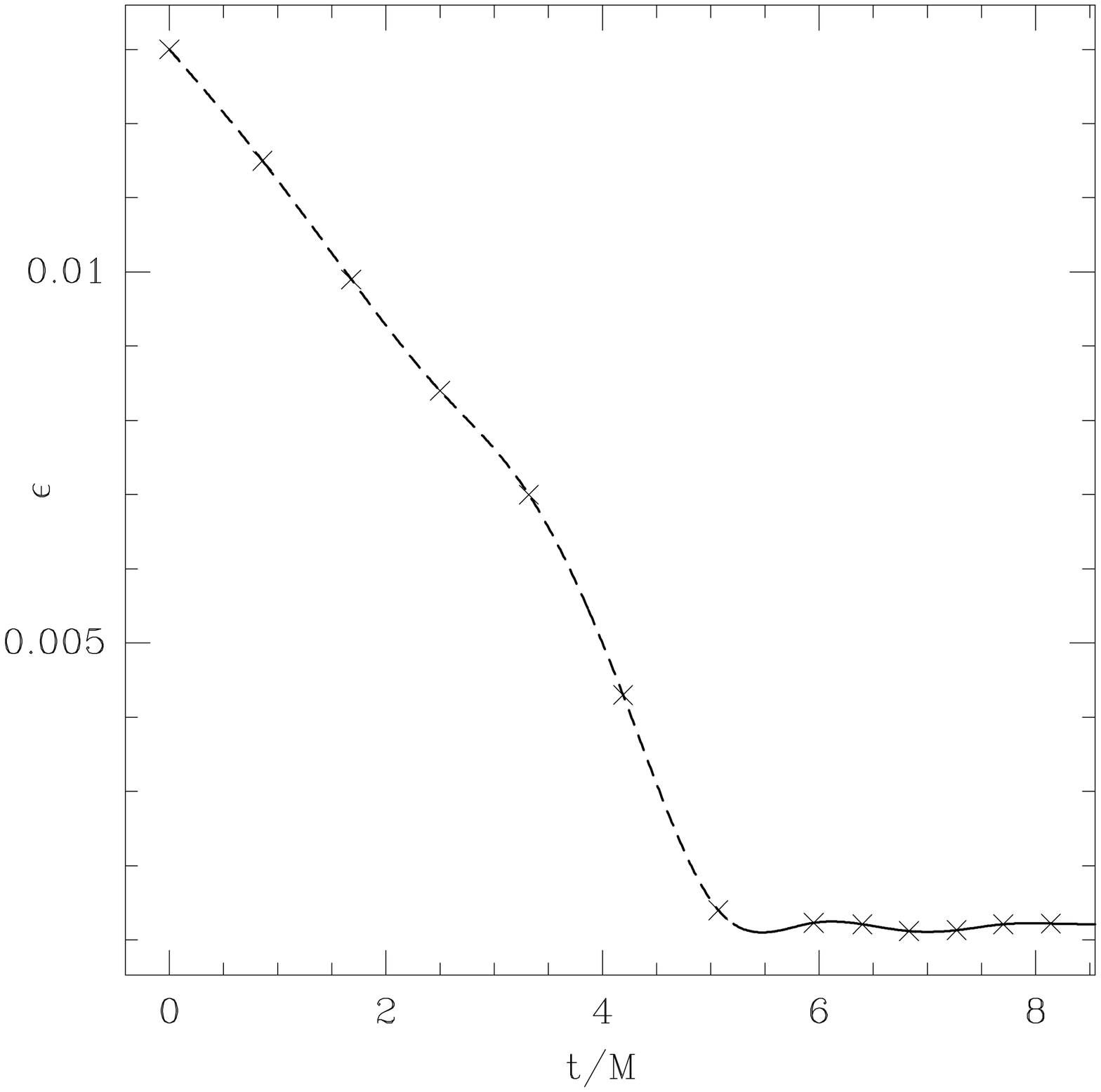}
\vspace{8.3cm}
\caption{
The radiated energy as a fraction of
the total mass-energy, determined with the perturbation
method, as a function of coordinate time
during a simulation of colliding clusters (Case b1).
The curve changes from dashed to solid
at the time an apparent horizon encompasses
both clusters.
Note that once this common horizon forms, the amount of radiation emitted
does not depend on when the perturbation calculation is performed.
}
\end{figure}
Typically, we wait until an
apparent horizon has formed and the distortions
of this horizon have become fairly small.
In the case of collisions, we wait until a common
apparent horizon, encompassing both objects, has formed.
We quantify the requirement of small distortions by requiring that the
proper polar and equatorial circumferences
of the apparent horizon be within $10\%$ of
their Schwarzschild values, i.e.,
$0.9 \lesssim C_{pol}/4 \pi M \lesssim 1.1$
and,
$0.9 \lesssim C_{eq}/4 \pi M \lesssim 1.1$.
We also require that the area of the apparent horizon
be within $5\%$ of its Schwarzschild value:
$0.95 \lesssim \sqrt{A^2/16\pi M^2} \lesssim 1.05$.

\begin{figure}
\label{fig:waveb1}
\special{hscale=0.45 vscale = 0.45 hoffset = -0.2 voffset = -4.
psfile=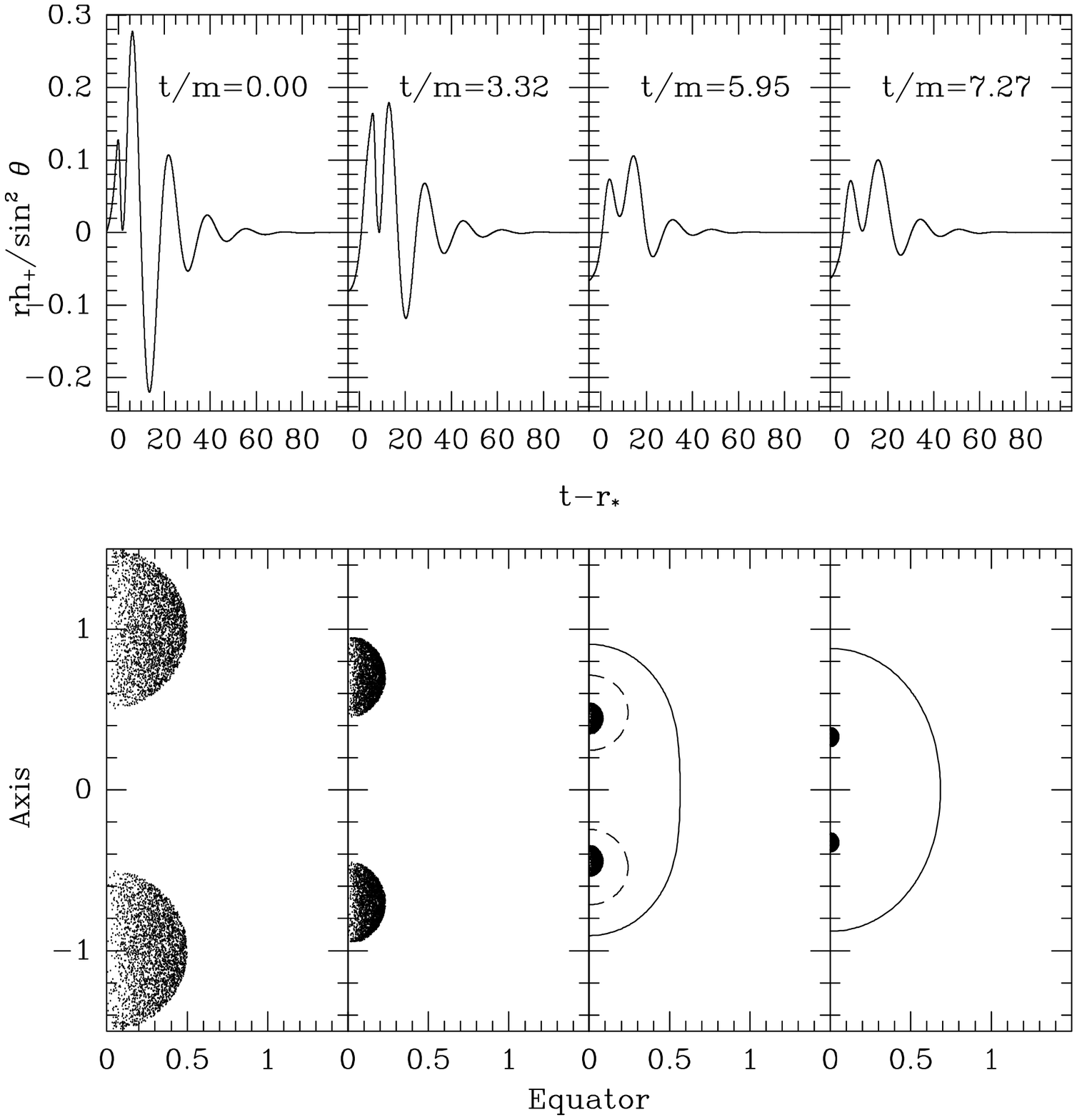}
\vspace{8.3cm}
\caption{
The perturbation method applied at different
times during a numerical simulation.   The lower
four frames show the particle positions for
a collision of initially  boosted star clusters
(Case b1).  The dashed lines indicate
disjoint marginally outer-trapped surfaces,
while the solid line shows the common apparent horizon.
The upper four frames (corresponding
to the same times) show the asymptotic waveform
computed using the perturbation method as
a function of retarded time in units of $M$.
Radial coordinates are also in units of $M$.
}
\end{figure}

Consider a concrete example that illustrates our approach.
In Fig.~\ref{fig:eoft} we show the radiated energy extracted when
using the perturbation method at different times during a
numerical simulation.  The black hole collision
shown, Case b1, is discussed below.
Fig.~\ref{fig:waveb1} shows the particle positions
and apparent horizon structure  at four times
during the course of the evolution.
The waveform extracted using the perturbation method
at each of those times is also shown.
After the common apparent horizon forms at a time
of about $t/M \simeq 5.0$, the waveform shape and
amplitude become fairly constant.  The
disjoint horizons form earlier, at $t/M \simeq 4.0$.
The efficiency -- the total radiated energy
divided by the total mass of the spacetime --
is shown in Fig~\ref{fig:eoft}.  This also levels off
to a nearly constant value once the black holes
are encompassed by a common horizon.
This effect is to be expected if the calculation is at all meaningful:
the final wave forms and energy emission should not depend on when the
numerical data is extracted for the perturbation equations once the
final black hole has formed and settled down.
Eventually, the effects
of throat stretching cause the numerical solution to become
inaccurate and the radiated energy begins to diverge
rapidly.  In all cases where the horizon settles
down to being nearly spherical
while the full numerical solution is still accurate
(judged by the constancy of the Brill mass~\cite{st92}),
the radiated energy remains constant as a function of time as required.

We did encounter some cases where is was impossible
to determine radiated energy to better than a factor
of 2.  Generally this was because the numerical
solution became inaccurate before the horizon settled
down.  These situations are exacerbated by the fact
that the quasi-isotropic spatial gauge conditions and
the Hamiltonian and momentum constraints solved in
our algorithm are elliptic equations. Hence, they
instantaneously propagate numerical
errors to the entire solution, even those occurring inside the
black hole horizon.  For this reason, even in the relatively accurate
cases, we focus
on the quadrupole, $\ell=2$ radiation.  At present
we do not have confidence in the lower amplitude $\ell=4$ wave forms as
the signals are comparable in size to the
numerical errors.

The table lists the 14 cases of colliding clusters and
disk collapses that were carefully studied using the
perturbation technique.  For each
case we give the parameters of the initial
data including proper separations where applicable.
The radiation efficiency is computed for each case.
In general, we were able to determine
the radiated energy to better than $20\%$ accuracy, and in many instances
to better than $< 5\%$.   In the remainder
of this section we detail these simulations
and discuss the emitted wave forms and radiated energies.

\subsection{Collisions of clusters and black holes}
\label{subsec:colresults}

An alternative to the vacuum, topological black hole approach
to simulating black hole collisions is the collision of black holes
that form from collapsing matter.
Here we consider nonequilbirium spheroids of collisionless matter that
separately undergo collapse before hitting each other head-on.
An advantage of
such an approach is that the topology is Euclidean so that
simple coordinate systems can be employed.  It is
also possible to study different scenarios including
those in which the clusters of matter do not individually
have marginally outer-trapped surfaces surrounding them
until after they are both inside a common event
(and possibly apparent) horizon.  The major
disadvantage of this approach is that the matter
source must be accurately tracked for the entire simulation --
even when it is inside the black hole.  This can
be very computationally intensive.  Of course, horizon
boundary conditions that would allow one to excise the
black hole interiors from the numerical simulation would
solve this problem.

In Sec.~\ref{sec:methods} we have briefly described the
fully general relativistic collisionless matter
evolution code used in these simulations.
Full details of this mean-field, particle simulation code, as well as
initial data and typical simulations of colliding clusters
can be found in Ref.~\cite{st92}.
All simulations discussed here were performed
on a 200 radial by 32 angular
zone grid with the matter source sampled with 3000 particles.
As mentioned above, we monitor the constancy of
quasi-local mass indicators such as the Brill mass for a check on
accuracy. For the
cases shown these resources were sufficient to keep the
Brill mass constant within a few percent.  This constancy
is crucial for the perturbation approach to give consistent
results.

First we consider initial data representing two clusters
of collisionless matter initially at rest.
Since this initial data is conformally flat and
at a moment of time-symmetry, it is only necessary
to specify the cluster radii and separation and
solve the Hamiltonian constraint to obtain
consistent data.  The analytic solution for the conformal
factor $\phi$ is given in Ref.~\cite{st92}. Interestingly,
in the vacuum exterior it agrees with the
solution given by Misner and Wheeler\cite{misner_wheeler57}
for two black holes in vacuum.  This black hole solution can be thought of
as having a a three-sheeted topology with two
Einstein-Rosen bridges each joining a separate
asymptotically-flat sheet to a single upper
asymptotically-flat sheet.  For cases
in which each cluster initially resides inside
its own apparent horizon one would expect identical
evolutions to result from the cluster initial
data and the 3-sheeted data.

Since most numerical studies(cf. Ref.~\cite{anninos_etal93})
of two black hole spacetimes have used two-sheeted
Misner data where an isometry is imposed between
the upper sheet and the single lower sheet, it
would be interesting to compare the radiation
expected from evolution of two-sheeted
initial data with the cluster data to determine
the importance of topology. Since both types of
data represent the same physical situation --
colliding black holes -- it would be disturbing
if the differences were large.
In a future study we will examine this issue in
more detail both for time-symmetric and boosted
initial data, which is difficult to implement numerically
in the 3-sheeted vacuum picture.

\begin{figure}
\label{fig:wavec0c5}
\special{hscale=0.45 vscale = 0.45 hoffset = -0.2 voffset = -4.
psfile=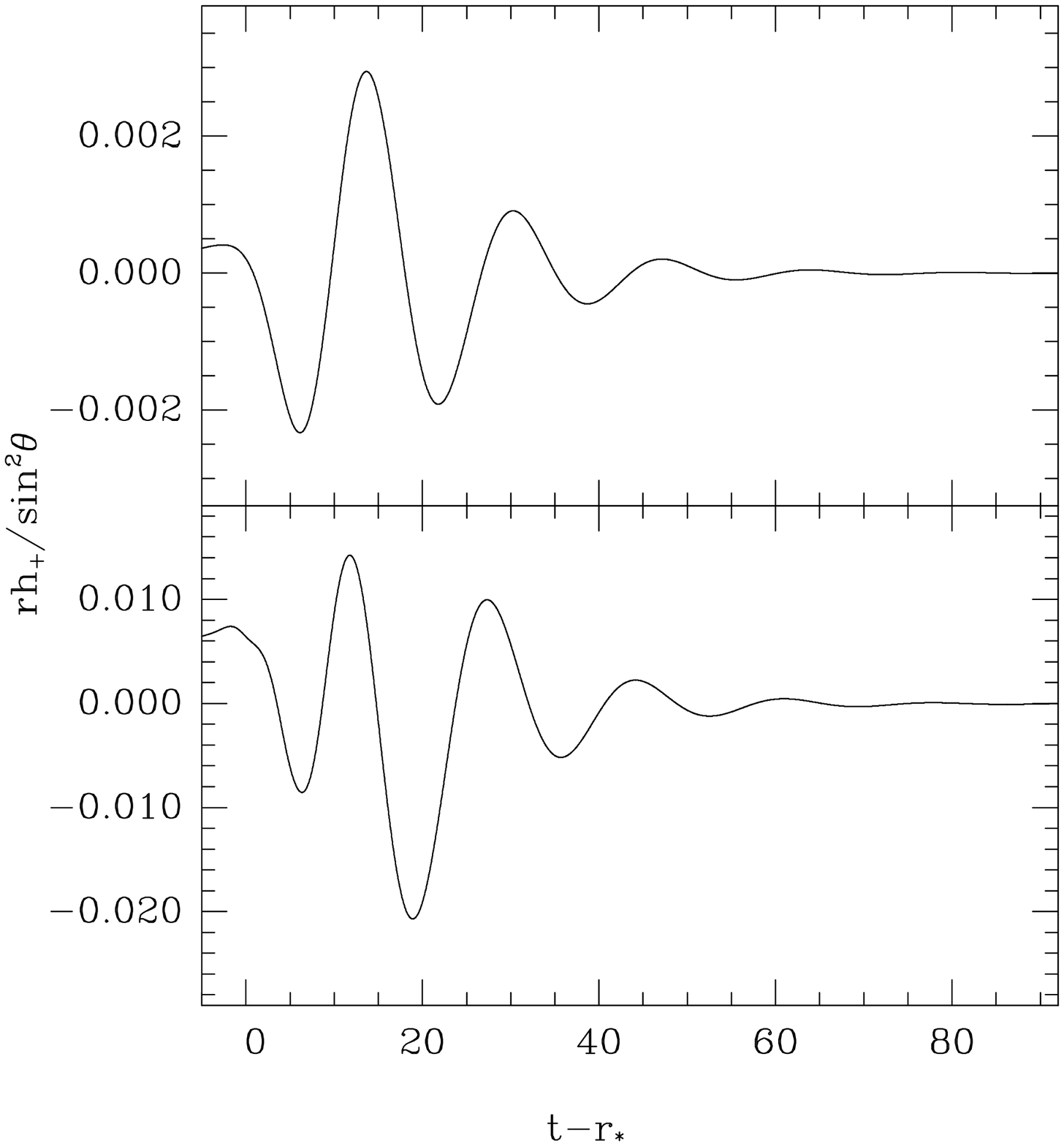}
\vspace{8.3cm}
\caption{
Wave forms from Case c0 (upper frame) and Case c5
(lower frame) are compared.  The asymptotic gravitational
wave amplitude computed with the perturbation method
is plotted as a function of retarded time in units of $M$.
}
\end{figure}

We have studied cluster collisions with a range of separations
and compactness.  In Case c0 the cluster centers
have an initial coordinate separation
of only $0.40M$ and there is a common apparent horizon
on the initial slice.
Not surprisingly, this case behaves like a single
perturbed black hole.  In the top frame of
Fig.~\ref{fig:wavec0c5} we show the waveform
from this case computed with the perturbation
method at an early time during the simulation.
The waveform is very reminiscent of
that found by Price and Pullin\cite{price_pullin94} for Misner
data in the small separation limit.
After a fairly short transient, the waveform is dominated
by the lowest order $\ell=2$ quasi-normal mode oscillations
with characteristic wavelength $\lambda = 16.8M$.

For comparison, in the same figure we show the waveform from
Case c5 in which the clusters initially have
a coordinate separation of $1.1M$. Here there
are initially no common or disjoint
apparent horizons.  The common horizon forms at a coordinate
time of about $2.0M$. The horizon is
initially very distorted, $C_{eq}\simeq 0.62$, $C_{pol}\simeq 1.16$.
The perturbation calculation of the waveform and
radiated energy is not carried out till $t \simeq 5.3M$
at which time the horizon has become nearly spherical.
Not surprisingly, the wave amplitude is much larger
in this case since the initial separation is
considerably larger. The waveform shapes are similar but not identical
because the cluster infall and collapse are
entirely hidden inside the black hole in Case c0.

We have also studied initial data,  Cases c2 and c4,
for which there are disjoint apparent horizons
surrounding each cluster.  These cases can
be thought of as true black hole collisions.
Since the clusters are more compact, the wave amplitudes in these
cases are considerably larger than those from cases
with the same {\it coordinate} separation but larger cluster radii.
For these cases it is possible to compute the
initial {\it proper} separation of the disjoint apparent horizons.
It turns out that the proper separation is indeed larger for these
compact cases. For these two cases
the radiated energy efficiencies
we obtain seem to be compatible with the results from
head-on collisions of time-symmetric topological
black holes\cite{anninos_etal93}.
Unfortunately, the throat stretching is very severe in these cases
and it is difficult to have confidence in
the energies and wave forms except for cases
with very small initial coordinate separations.

We have also considered cases for which the clusters
are boosted towards each other, like those evolved in Ref.~\cite{st92}.
Since these data sets are no longer
time-symmetric, it is necessary to solve both the Hamiltonian
and momentum constraints for the initial gravitational
field.  These initial value equations are solved by iteration as described
in Ref.~\cite{st92}.
The advantage of the boosted calculations
is that the clusters can be started at larger
coordinate separations and still collide before the
numerical results become unreliable because of the intrinsic collapse
of each cluster.  As can be
seen in the table, the boosted Cases b1-b4 produce
considerably more radiation than the unboosted clusters. With boosts, the
empirical limit
on radiation efficiency of $\sim 10^{-3}$ for
black hole collisions resulting from
time-symmetric data (cf. Ref.~\cite{abrahams_cook94})
can be exceeded.
The largest amount of radiation, efficiency $5\times 10^{-3}$,
is produced in Case b2,
which has the largest separation and boost velocity.

\begin{figure}
\label{fig:waveb3b4}
\special{hscale=0.45 vscale = 0.45 hoffset = -0.2 voffset = -4.
psfile=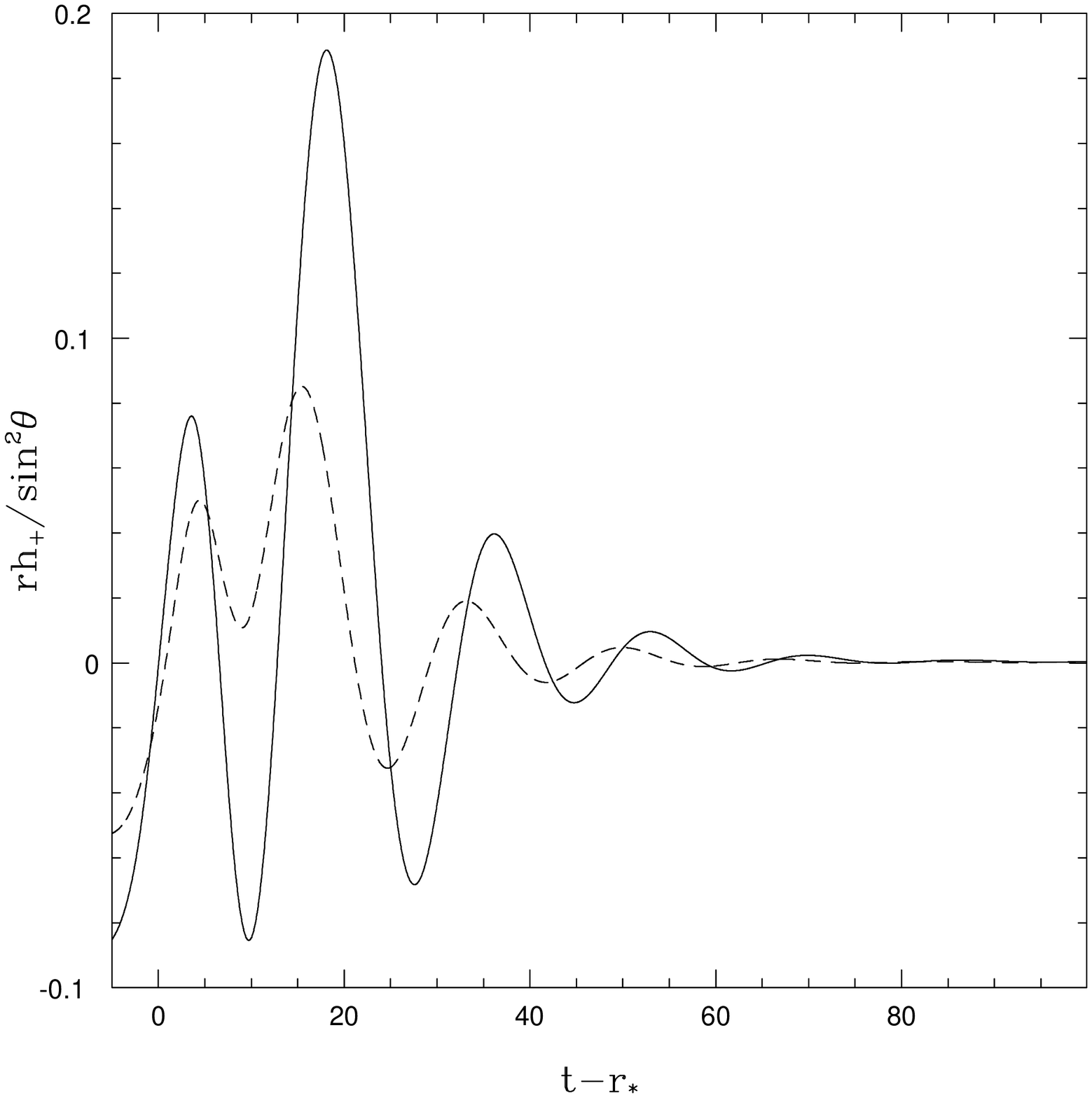}
\vspace{8.3cm}
\caption{
The gravitational waveform from Case b3 (dashed curve) is compared
with the waveform from Case b4 (solid curve).  The
asymptotic wave amplitude calculated with the perturbation
method is plotted as a function of retarded time in units
of $M$.
}
\end{figure}

Fig.~\ref{fig:waveb3b4} compares wave forms from two
boosted clusters, Cases b3 and b4.  The coordinate separation and
initial radii of the clusters
is the same in each case, only the boost velocity is different.
Clearly, the wave amplitude is much greater in the
case of the more violent collision, Case b4.
The wave forms from the boosted cases all
have the same basic qualitative appearance.  The initial
hump, not present in analysis of boosted Misner-like
data~\cite{abrahams_cook94}, probably arises because
the clusters are not initially black holes.
Quasi-normal modes dominate
after about $20M$ in all cases.

\subsection{Disk collapse}
\label{subsec:diskresults}

Collapsing disks of collisionless matter provide an
excellent test problem for axisymmetric numerical relativity
codes.  Since the source is all in the equatorial
plane, the matter evolution equations are all
one-dimensional.  The source is felt by the
two-dimensional gravitational field via
jump conditions.  When the disk matter is initially
at rest, the situation provides an interesting analogy
to Oppenheimer-Snyder collapse to
a black hole in spherical symmetry, but with
the added important feature of gravitational
radiation production.
Since the gravitational field is dynamical in disk collapse, the
full machinery of numerical relativity is required to follow the evolution,
while spherical Oppenheimer-Snyder collapse is analytic.
In Ref.~\cite{ast94} the basic equations for
this model were given and a number of test simulations
discussed.  Here we restrict our attention to cold matter
cases that collapse to black holes.
By cold matter we mean cases in which the particles in the disk have
no velocity dispersion at $t=0$.
All simulations shown here were performed on a
300 radial by 16 angular
zone grid with the matter source represented
by 12000 particles. Even though disk collapse is in a sense the simplest
radiating system, in no case are we able to track the evolution long
enough to read off the full waveform directly from the simulation.
Only with the perturbation method can we extract the entire waveform.

\begin{figure}
\label{fig:d3part}
\special{hscale=0.45 vscale = 0.45 hoffset = -0.2 voffset = -4.
psfile=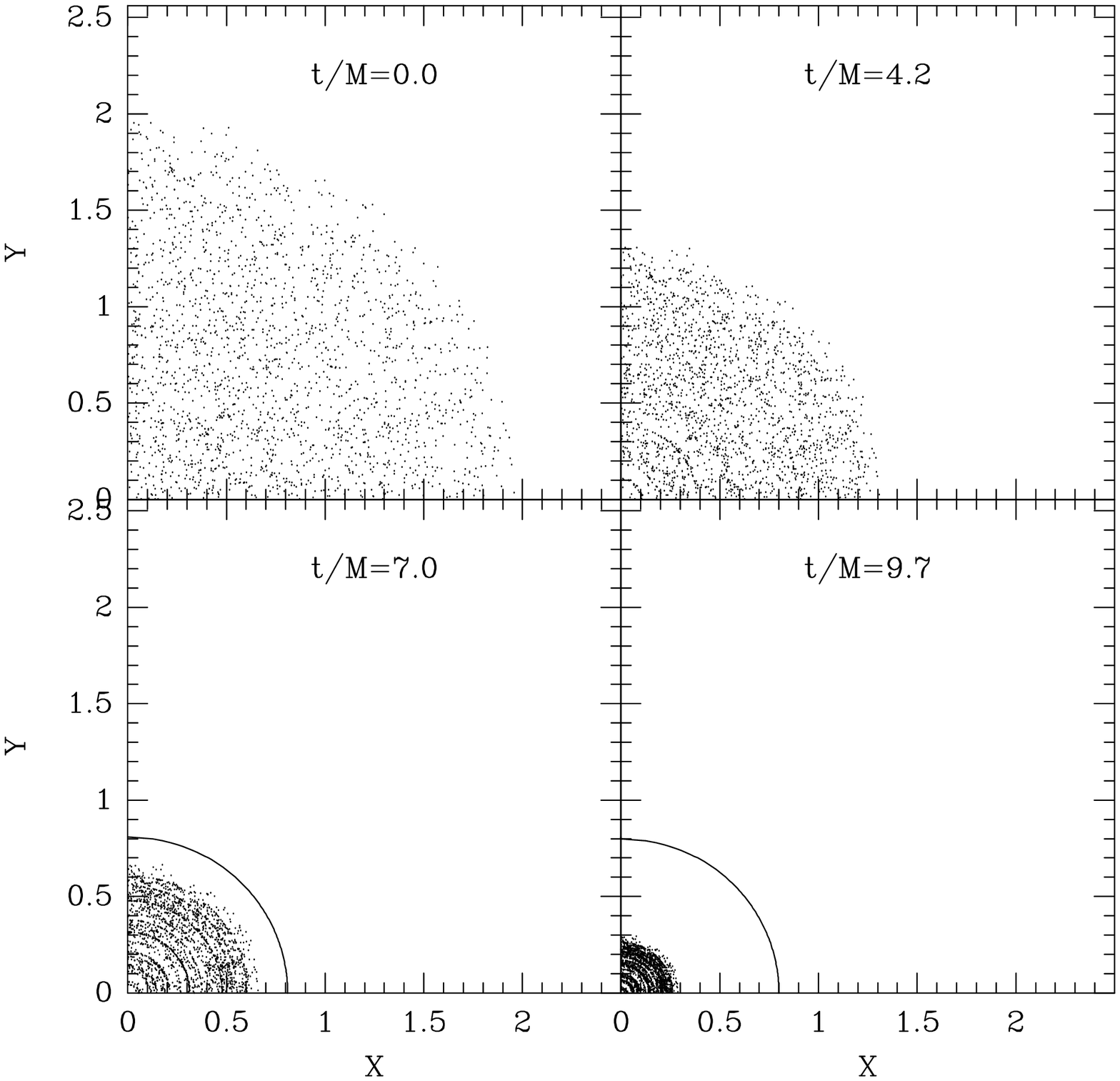}
\vspace{8.3cm}
\caption{
Particle positions in the equatorial plane are plotted
at several times during a disk collapse, Case d3.
The apparent horizon is indicated by the solid line
in the third and fourth frames.
}
\end{figure}

\begin{figure}
\label{fig:d3nmfit}
\special{hscale=0.45 vscale = 0.45 hoffset = -0.2 voffset = -4.
psfile=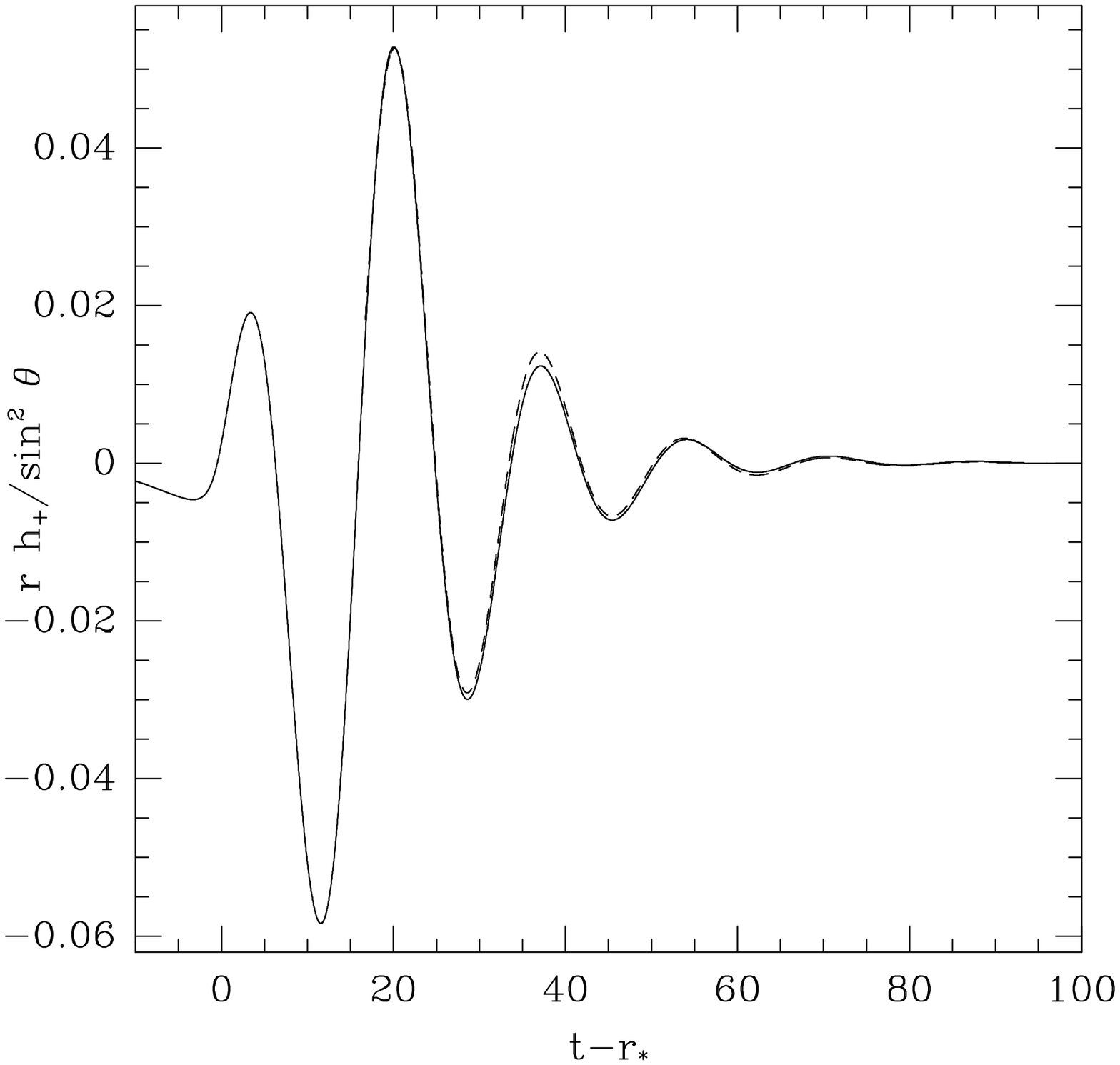}
\vspace{8.3cm}
\caption{
The gravitational waveform is plotted as a function
of retarded time in units of $M$
for disk collapse Case d3 (solid line).  The least-squares
fit to the two lowest-order $\ell=2$ Schwarzschild
quasi-normal modes is also shown (dashed line) for
$t-r_* > 17M$.
}
\end{figure}

We consider three disk collapses, Cases d1-d3
in the table.  As might be expected the larger
the initial radius, the more radiation is
produced, although the efficiencies are very small,
less than $0.1\%$ in all cases.
Fig.~\ref{fig:d3part} shows a typical collapse,
Case d3.  The matter starts at rest with initial
disk radius $R/M=2.0$.
The locations of the particles in the equatorial
plane are shown.  After the matter
passes inside a radius of about $r/M \simeq 0.5$
an apparent horizon forms.
As in Newtonian gravitation, cold disks are unstable to ring formation.
As Ref.~\cite{st92} shows, ring formation here
mostly occurs after the black hole forms.
The waveform computed using the perturbation approach
at a time of about $t/M=8$ is shown in
Fig.~\ref{fig:d3nmfit}.
A least squares fit to
the two lowest-order $\ell=2$ Schwarzschild quasi-normal
modes is also shown. After an initial transient
feature, lasting for about $15M$, the fit is excellent
in both amplitude and frequency.

\begin{figure}
\label{fig:d2comp}
\special{hscale=0.45 vscale = 0.45 hoffset = -0.2 voffset = -4.
psfile=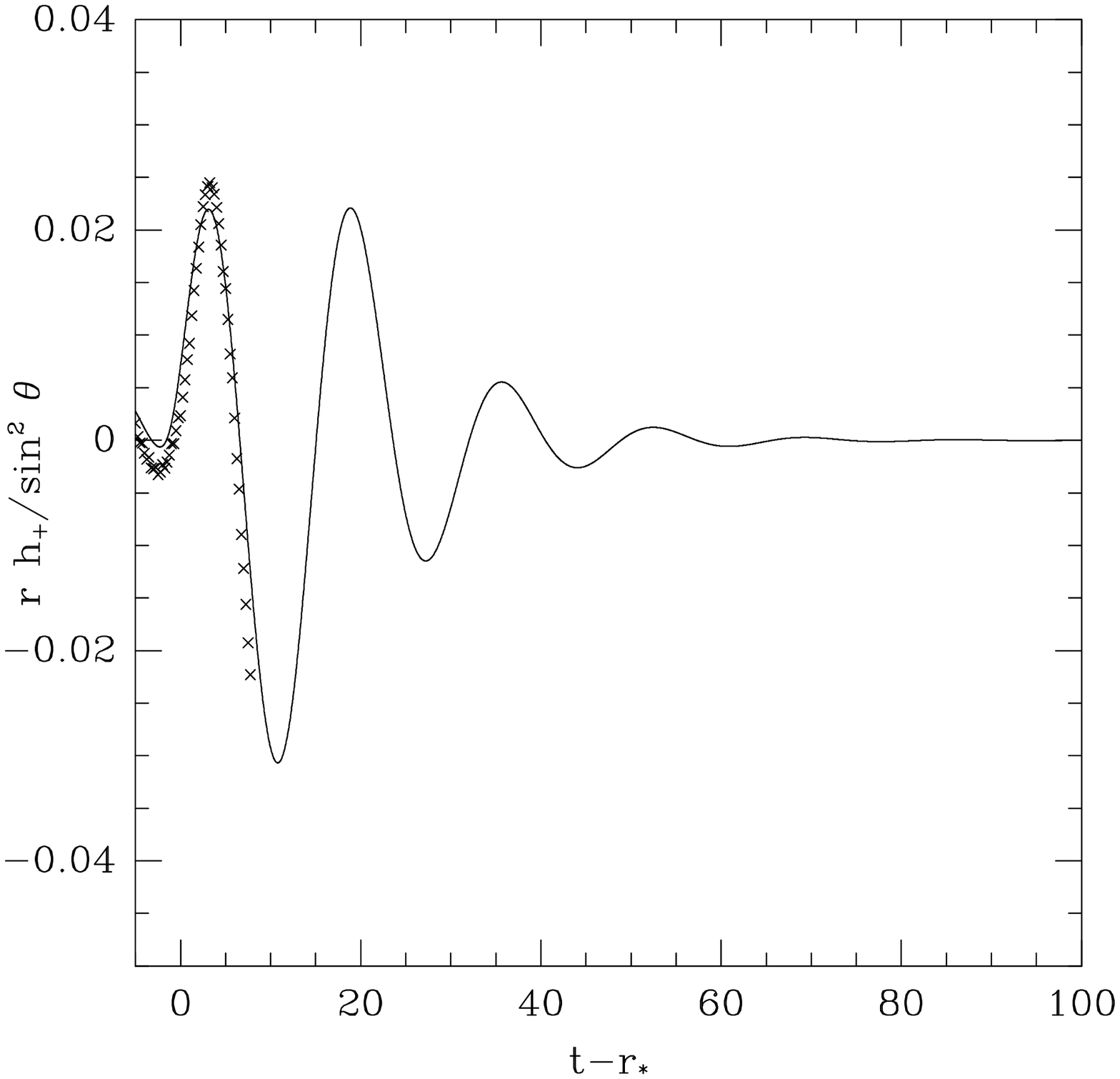}
\vspace{8.3cm}
\caption{
The gravitational waveform computed using the perturbation
method for disk collapse Case d2 is plotted as a
function of retarded time in units of $M$.  For comparison, the
gravitational waveform computed at a radius $r/M=8$
using the standard spacelike radiation extraction technique
is also shown (points).
}
\end{figure}

In Fig.~\ref{fig:d2comp} we compare the
%
waveform computed with the perturbation technique
(using simulation data at
a time of about $t=4.7M$) with that extracted
using the usual technique \cite{ae90}
for Case d2 (see Ref.~\cite{ast94}).  Note
that the simulation terminated within about $15M$ after black hole
formation, leaving little time for waves to reach the numerical
detector at $r=8M$. Only the radiation from
the infall phase and the first oscillation of the black
hole are obtainable using the standard method
before the simulation goes bad.
The agreement between the two methods up to this time is reasonable considering
that no special care has been taken in separating
out the near-zone field for the infall part of the
waveform.  This is a delicate matter requiring special
determination of initial values for the spacelike
extraction techniques~\cite{abrahams92}, but is automatically
taken care of in the perturbation approach provided that the
waveform is read off the Zerilli integration at
a sufficiently large radius.

\section{Discussion}
In summary, we have found that the perturbation
approach to calculating
asymptotic wave forms and radiated energies is effective not only in analyzing
initial
data sets, but can also be used in conjunction with
numerical relativity simulations.  For
many scenarios it is difficult to evolve black holes
long enough to compute accurately the wave forms
produced by their formation and ring-down. As long as the simulation
can be prolonged to the point that a fairly spherical
black hole horizon can be identified, the perturbation method presented here
should give reliable results. The method holds great promise for
more general systems, since it should be possible to extend the method
to perturbations of a rotating black hole.
\acknowledgments
We thank R.~H. Price for stimulating conversations.
This work was supported by National Science Foundation
grants AST 91-19475 and PHY 94-08378 and
the Grand Challenge grant NSF PHY 93-18152 / ASC 93-18152
(ARPA supplemented).  Computations
were performed at the Cornell Center for Theory and
Simulation in Science and Engineering, which is supported
in part by the National Science Foundation, IBM Corporation,
New York State, and the Cornell Research Institute.
\newpage

%
%
%
%

%
%
\newpage
\twocolumn
\widetext
\begin{table}
\caption{Radiated energy efficiencies from cluster collisions
and cold disk collapse.}
\label{table:col}
\begin{tabular}{l|c|c|c|c|c|c|c}
Case&description&$z_0/M$
\tablenote{Coordinate displacement up the axis of the
cluster center.}
&$R_0/M$ \tablenote{Coordinate radius of cluster or disk.}&$L/M$
\tablenote{Proper separation of cluster centers.}&$L_{ah}/M$
\tablenote{Proper separation of disjoint apparent horizons.}&$v/c$
\tablenote{Boost velocity of cluster as measured by a normal
observer.}&$\epsilon$
\tablenote{The radiation efficiency.}\\
\hline \\
c0&collision&0.20&0.15&6.97&&0.0&$1.5\times10^{-6}$ \\
c1&collision&0.40&0.3&5.1&&0.0&$1.5\times10^{-5}$ \\
c2&collision&0.40&0.1&9.6&3.1&0.0&$1.4 \times 10^{-4}$ \\
c3&collision&0.475&0.3&5.7&&0.0&$5.5\times10^{-5}$ \\
c4&collision&0.50&0.1&10.2&3.6&0.0&$3.\times10^{-4}$ \\
c5&collision&0.55&0.3&6.1&&0.0&$7.5\times10^{-5}$ \\
c6&collision&0.65&0.3&6.4&&0.0&$9.0\times10^{-5}$ \\
b1&boosted collision&1.0&0.5&6.2&&0.05&$1.2\times10^{-3}$ \\
b2&boosted collision&1.4&0.5&6.5&&0.15&$5.\times10^{-3}$\\
b3&boosted collision&0.75&0.5&5.3&&0.05&$7.\times10^{-4}$\\
b4&boosted collision&0.75&0.5&5.1&&0.10&$4.\times10^{-3}$ \\
d1&disk&&$1.00$&&&&$1.5\times10^{-4}$ \\
d2&disk&&$1.50$&&&&$3.3\times10^{-4}$ \\
d3&disk&&$2.00$&&&&$5.3\times10^{-4}$ \\
\hline
\end{tabular}
\end{table}

\end{document}